\documentclass[conference,a4paper,10pt]{IEEEtran}
\usepackage{amsfonts}
\IEEEoverridecommandlockouts
\usepackage{epsfig}
\usepackage{graphicx}
\usepackage{psfig}
\usepackage{subfigure}
\usepackage{epsf}
\usepackage{epstopdf}
\usepackage{bbm}
\usepackage{filecontents}
\usepackage[noadjust]{cite}
\usepackage{amsmath}
\allowdisplaybreaks[4]
\usepackage{booktabs}
\usepackage{fancyhdr}

\usepackage{algorithm}
\usepackage{algorithmic}
\usepackage{epstopdf, cite,color}
\usepackage{amsthm}
\usepackage{amssymb}
\newcommand{\bc}{\begin{center}}
	\newcommand{\ec}{\end{center}}
\newcommand{\be}{\begin{equation}}
\newcommand{\ee}{\end{equation}}
\newcommand{\bea}{\begin{eqnarray}}
\newcommand{\eea}{\end{eqnarray}}

\usepackage[letterpaper, left=0.625in, right=0.6in, bottom=1.03in, top=0.7in]{geometry}

\begin{document}

\title{Efficient Phishing URL Detection Using Graph-based Machine Learning and Loopy Belief Propagation}

\author{Wenye Guo\textsuperscript{\S}, Qun Wang\textsuperscript{\S}, Hao Yue\textsuperscript{\pounds}, 
     Haijian Sun\textsuperscript{\dag}, and Rose~Qingyang~Hu\textsuperscript{\ddag}\\
      \textsuperscript{\S}Department of Computer Science, San Francisco State University, San Francisco, CA, 94132\\
     
     % $^*$Jiangsu Key Laboratory of Wireless Communications, \\Nanjing University of Posts and Telecommunications, Nanjing, China, 210000\\
     
     \textsuperscript{\pounds}Department of Computer Science and Engineering, University of California, Santa Cruz, CA, 95064\\
     
     \textsuperscript{\dag}School of Electrical and Computer Engineering, University of Georgia, GA, 30602\\
     
     \textsuperscript{\ddag}Department of Electrical and Computer Engineering, Utah State University, Logan, UT, 84341.\\
   Emails:  
wenyeguo@outlook.com, claudqunwang@ieee.org, hyue11@ucsc.edu,
\\hsun@uga.edu, rose.hu@usu.edu}
\maketitle	

\IEEEpeerreviewmaketitle
\begin{abstract}
%In today's digital age, with the widespread adoption of mobile devices, the Internet has become essential to our daily lives. However, it also exposes us to a range of cyber threats, ranging from malware dissemination to phishing attacks. Malicious URLs serve as conduits for various cybercriminal activities, from malware dissemination to phishing attacks. Addressing this issue requires robust detection methods to identify and mitigate such threats effectively. To tackle this challenge, we present a novel graph-based machine-learning approach for identifying malicious URLs. By leveraging Loopy Belief Propagation (LBP) with an efficient convergence strategy, we have significantly enhanced the method's performance. Our experiments conducted on diverse datasets demonstrate remarkable results, with an F1 score of up to 98\%. This method not only improves detection accuracy but also offers reliability and reproducibility, making it a valuable contribution to the field of cybersecurity.

The proliferation of mobile devices and online interactions have been threatened by different cyberattacks, where phishing attacks and malicious Uniform Resource Locators (URLs) pose significant risks to user security. Traditional phishing URL detection methods primarily rely on URL string-based features, which attackers often manipulate to evade detection. To address these limitations, we propose a novel graph-based machine learning model for phishing URL detection, integrating both URL structure and network-level features such as IP addresses and authoritative name servers. Our approach leverages Loopy Belief Propagation (LBP) with an enhanced convergence strategy to enable effective message passing and stable classification in the presence of complex graph structures. Additionally, we introduce a refined edge potential mechanism that dynamically adapts based on entity similarity and label relationships to further improve classification accuracy. Comprehensive experiments on real-world datasets demonstrate our model's effectiveness by achieving  F1 score of up to 98.77\%. This robust and reproducible method advances phishing detection capabilities, offering enhanced reliability and valuable insights in the field of cybersecurity.

\end{abstract}
	
\begin{IEEEkeywords}
Phishing URL Detection, Graph-based Models, Loopy Belief Propagation, Feature Augmentation, Network-based Detection
\end{IEEEkeywords}
\IEEEpeerreviewmaketitle
\section{Introduction}
% In today’s digital landscape, cyber threats such as malware dissemination, spamming, and phishing attacks pose significant risks to individuals and organizations. Among these, malicious websites are a primary channel through which cybercriminals execute attacks. Phishing websites, in particular, impersonate legitimate businesses, financial institutions, or trusted organizations to deceive users and steal sensitive information. Early detection of phishing URLs is, therefore, critical to mitigate these threats and protect users.

In today’s digital landscape, cyber threats such as malware dissemination, spam, and phishing attacks pose significant risks to individuals and organizations. Malicious websites are a primary channel for these attacks, with phishing sites impersonating legitimate businesses, financial institutions, or trusted entities to deceive users and steal sensitive information \cite{intro1}. Beyond traditional web applications, URLs are also extensively used in emails and text messages, where they often mimic legitimate URL patterns to mislead recipients, redirecting them to malicious servers. These phishing URLs can lead to privacy breaches and password theft and even trigger further attacks involving trojans and backdoors \cite{intro2}. Accurate and early detection of phishing URLs is therefore essential to mitigate these risks and protect users\cite{intro3}.

Different machine-learning approaches have been proposed for phishing detection.
A deep transformer model pre-trained on over 3 billion unlabeled URLs with a tailored objective for phishing detection was proposed in \cite{intro2}, achieving superior efficiency and robustness through supervised fine-tuning with adversarial methods.
In \cite{rl1}, a deep learning-based phishing detection method has been proposed by analyzing login page URLs and achieving 96.50\% accuracy.
In \cite{rl2},  the author developed an automated machine learning framework for real-time phishing URL detection, achieving 87\% accuracy. The authors of \cite{rl3} introduced a transformer-based model that demonstrates robustness against adversarial attacks. 
The authors in\cite{ref} introduced a network-based inference method to detect phishing URLs designed to evade detection by mimicking legitimate patterns and achieving higher robustness.

Most of the existing works rely heavily on URL string-based features, such as specific words, punctuation, or URL length. However, attackers frequently manipulate these features to evade detection, making traditional methods less effective.
To overcome these limitations, we incorporate traditional URL-based features with IP addresses and authoritative name servers, which are more stable and less prone to manipulation. IP addresses represent the server's numerical location and are typically more difficult to alter, while authoritative name servers manage the translation of domain names into IP addresses, providing additional stability.
Based on this, we propose a novel graph-based machine learning model that integrates network-based Loopy Belief Propagation (LBP) with an enhanced convergence technique. Our approach is designed to improve phishing detection by leveraging both URL structure and network-level features. 
%Through extensive testing on two real-world datasets, our model consistently achieves high F1 scores of up to 98\%, demonstrating its reliability and effectiveness in detecting phishing URLs.
Our contributions are fourfold:
 (1) By integrating IP addresses and authoritative name servers alongside traditional URL features, we enhance the model’s resilience to adversarial manipulation. 
 %These network-level features, unlike URL strings, are more challenging for attackers to modify, providing additional stability in phishing detection.
 (2) Moreover, We introduce a refined edge potential mechanism within LBP that adapts dynamically based on the similarity and label relationships of connected entities. %This method improves classification accuracy by making the edge potentials more context-aware, reducing false positives in phishing detection.
(3) In addition, to optimize performance, we enhance the convergence technique in LBP to enable faster and more stable message passing. 
(4) We conducted extensive testing on both established and newly collected datasets, increasing the dataset’s diversity and size. This expansion helps the model generalize better across different phishing scenarios, resulting in consistent F1 scores of up to 98.77\%, demonstrating the model’s robustness and reliability.

The subsequent sections of this paper are organized in the following manner. The system model is presented in Section II. The proposed approach is developed in Section III. The findings of the simulation are presented in Section IV. Finally, Section V provides the concluding remarks for this paper.
%我们的贡献有4，我们结合了ip和dns，我们提出了新的edge potential计算方法，我们改进了收敛速度，我们增加了数据集并提升了检测精度

\section{System Model}
%In this section, we will begin by introduce our proposed phishing detection framework. Then delve into the concept of message passing, serving as the foundation for label inference, and elaborate on the utilization of Loopy Belief Propagation (LBP) for phishing node classification. Lastly, we will outline our algorithmic process in detail.
\begin{figure}
\setlength{\abovecaptionskip}{-0.2cm} 
	 \setlength{\belowcaptionskip}{-1cm}
\centering
\includegraphics[width=0.45 \textwidth]{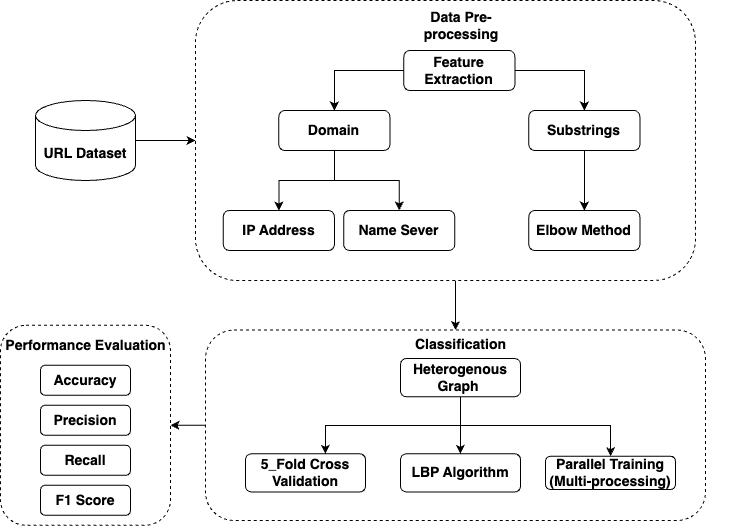}
\caption{The overall workflow of the proposed method. \label{fig1}}
\end{figure}
 The proposed phishing detection framework is shown in Fig.\ref{fig1}.  We first pre-process the dataset and extract features from the URL datasets. Subsequently, we construct a heterogeneous graph based on both the URLs and the extracted features. Finally, we employ a 5-fold cross-validation process, which is executed in parallel to train our model. Following model training, we evaluate its performance.
 %using the F1 score. Fig.\ref{fig:overview} depicts a summary of the methodological steps of our work, which will be elaborated upon in detail in the subsequent subsections.\par

\subsection{Data Processing and Feature Extraction}
We first extract features from both URL components and network-based information to enhance phishing detection accuracy. Unlike previous methods that primarily rely on URL string-based features \cite{baseline1, baseline2,baseline3, baseline4, baseline5}, our approach incorporates both IP addresses and authoritative name servers, which are less prone to manipulation by attackers.
%as shown in Fig.~\ref{fig:data_preprocessing}.

\begin{figure}
\setlength{\abovecaptionskip}{-0.2cm} 
	 \setlength{\belowcaptionskip}{-1cm}
\includegraphics[width=0.45\textwidth]{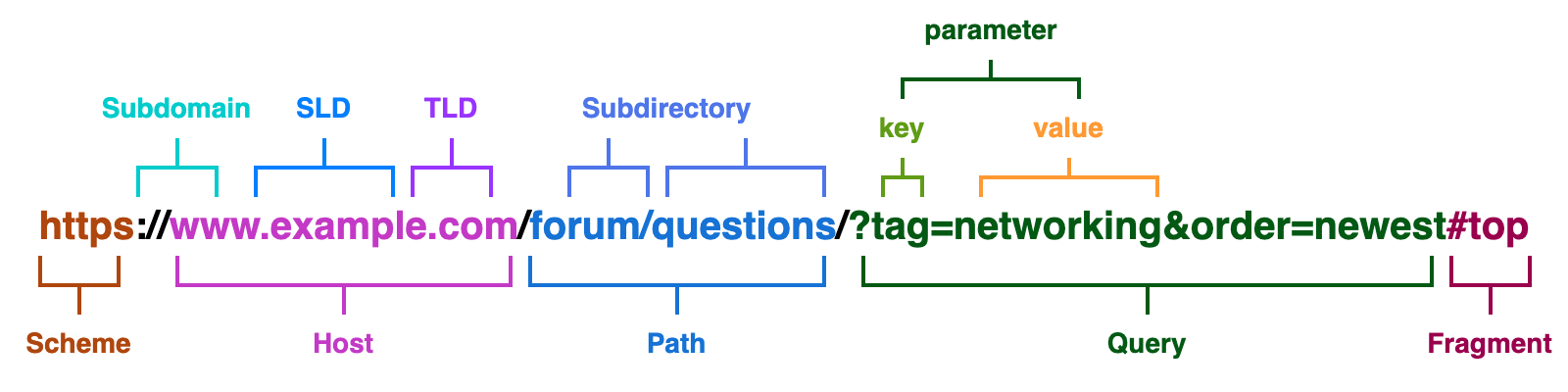}
\caption{URL anatomy diagram.}
\label{fig:url}
\end{figure}

% \begin{figure}
% \setlength{\abovecaptionskip}{-0.2cm} 
% 	 \setlength{\belowcaptionskip}{-1cm}
% \centering
% \includegraphics[width=0.45\textwidth]{figures/data_preprocessing.png}
% \caption{Feature Extraction}
% \label{fig:data_preprocessing}
% \end{figure}
The URL anatomy consists of scheme, subdomain, top-level domain, second-level domain, subdirectory, parameter, path, query, and fragment, as shown in Fig.~\ref{fig:url}. We split URLs into several segments, each of which contributes to feature extraction. 
The following are key components used in our model: 
%\begin{itemize} 
(1) Domain and Subdomains: Extracted and analyzed for common phishing patterns, separating subdomains, second-level domains (SLDs), and top-level domains (TLDs). 
(2) Path and Queries: Segmented based on punctuation and special characters to capture embedded malicious commands or obfuscated URLs. 
(3) IP Addresses and Nameservers: These are included as more stable features, as these are less frequently altered compared to URL strings, enhancing resilience against evasion tactics. 
%\end{itemize}
This comprehensive feature set enables our graph-based model to more effectively identify phishing attempts by leveraging both structural and network-level indicators.
% \begin{itemize}
%     \item The scheme specifies the protocol used to access a web page, such as ``http'' or ``https''. However, due to their high frequency, we do not utilize schemes as features in our analysis.
%     \item The host typically contains the domain name or IP address. As illustrated in Fig.~\ref{fig:url}, the domain name consists of three main parts: subdomains, the second-level domain (SLD), and the top-level domain (TLD). These parts are separated by symbol ``.''.
%     \item Occasionally, prior to the host, a URL may include a username and password, structured as follows:    ``http://username:password@example.com''. If the URL adheres to this format, we segment the URL using punctuation symbols ``//'', ``:'', and ``@''.
%     \item The path is used to specify the location of resources on the server. To segment the path, we utilize all possible punctuation symbols, including ``/'',``.'',``!'',``\&'',``,'',``\#'', ``$\backslash$'', ``\$'', ``\%'', ``;''.
%     \item A query string, an optional URL component, is preceded by a question mark and may contain multiple queries separated by ``\&". Each query comprises a key-value pair connected by ``=''. We extract substrings from the query string using ``\&'' and ``='' symbols.
% \end{itemize}

\begin{figure}
\setlength{\abovecaptionskip}{-0.2cm} 
	 \setlength{\belowcaptionskip}{-1cm}
\centering
\includegraphics[width=0.45\textwidth]{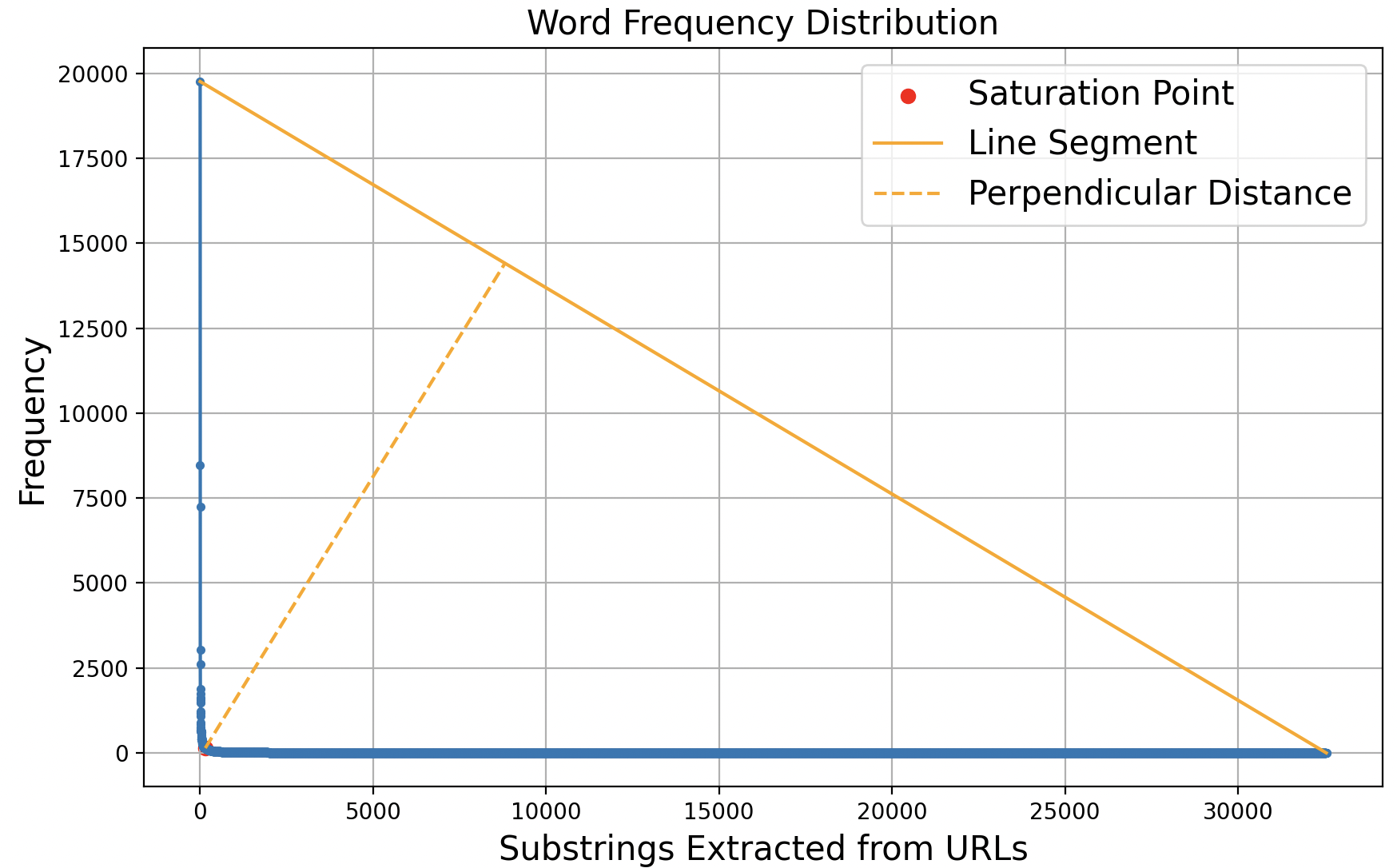}
\caption{Elbow method saturation point plot.}
\label{fig:elbow}
\end{figure}

% Our extracted substrings (as depicted in Fig.~\ref{fig:elbow}) exhibit a distribution similar to the findings of \cite{ref}, which indicated consistency with Zipf’s law among the extracted words from URLs. Therefore, we employed the elbow method \cite{elbow} to remove the high-frequency, meaningless words. 

% The elbow method identifies the saturation point, which is determined by finding the point where the perpendicular distance to the line segment connecting the two ends of the curve is the largest (as shown by the red point in Fig.\ref{fig:elbow}). Consequently, we remove all words from the substrings whose frequency values exceed this point.

The distribution of extracted substrings follows a pattern consistent with Zipf’s law, as observed in previous studies \cite{ref}. To eliminate high-frequency, low-value words, we applied the elbow method \cite{elbow}, which identifies the optimal cutoff point for feature selection. As shown in Fig.\ref{fig:elbow}, the saturation point is marked by the maximum perpendicular distance from the curve to the line segment and is used to filter out substrings with frequencies above this threshold.

\subsection{DNS and IP Address-based Feature Enhancement}

%To extract authoritative nameservers and IP address associated with a URL, we begin by isolating the domain of the URL, which typically comprises the second-level domain (SLD) and the top-level domain (TLD). For example, in the URL ``http://www.example.com'', the domain name is "example.com".Once we have obtained the domain, we execute Domain Name System (DNS) queries to retrieve the NS records associated with it, thereby identifying its authoritative nameservers. In our implementation, we obtain nameservers using a command like ``dig +short NS domain\_name'', where ``domain\_name'' is the domain name we want to query. This command specifically requests the NS (Name Server) records for the specified domain, which typically represent the authoritative nameservers responsible for the domain. Additionally, we utilize VirusTotal API, as recommended by \cite{ref}, to obtain its IP address.

To extract authoritative nameservers and IP addresses for a given URL, we first isolate the domain, typically composed of the SLD and TLD (e.g., "example.com" from http://www.example.com). We then use DNS queries to retrieve the nameserver (NS) records, identifying the authoritative nameservers. In our implementation, this is achieved using commands like \textit{dig +short NS domain\_name} to query the NS records. Additionally, we use the VirusTotal API to obtain the associated IP addresses, as recommended by \cite{ref}.

\section{Graph-based Machine Learning Model}
%A URL serves as a human-readable format that indicates the resource's location on the internet.
Attackers often manipulate URL strings to make phishing sites appear legitimate. However, modifying nameservers and IP addresses is more difficult. 
Phishing URLs often exhibit similar string patterns and may utilize the same IP addresses, indicating interconnected relationships among entities. Therefore, employing a graph-based model enables us to detect phishing URLs by analyzing the connections between URLs and features.

\subsection{Graph Construction}
As illustrated in Fig.~\ref{fig1}, after extracting all the necessary features, we construct the graph with the extracted features. We establish edges between the following entities:
(1) URL and its corresponding domain.
(2) Domain and its associated IP addresses.
(3) Domain and its authoritative nameservers.
(4) URL and its sub-strings.
Each graph node contains metadata such as its actual label and the prior probability, which is either inferred from a traditional machine learning model or set to a default value. For more information, please refer to Table~\ref{table:graph}.

\subsection{Model Operation and Inference}
%We employ loopy belief propagation (LBP) \cite{lbp} alongside an advanced edge potential assignment mechanism, which incorporates a penalty to enhance performance, following the approach outlined in \cite{ref}. Our primary contribution in this step involves introducing a novel convergence approach to enhance the convergence rate, thereby improving performance and reliability. We will elaborate on this in Sec.\ref{sec:convergence}.\par
%
We employ LBP with an enhanced edge potential assignment mechanism by incorporating a penalty to improve performance \cite{ref} \cite{lbp} . Our primary contribution is a novel convergence strategy that accelerates convergence and enhances reliability. We will elaborate on this in Sec.\ref{sec:convergence}.

\begin{table}[h]
\vspace{-0.2cm}
\centering
\vspace{-0.2cm}
\caption{The information of graph node. \label{table:graph}}
\begin{tabular}{|p{2cm}|p{6cm}|}
\hline
\textbf{Name}  &\textbf{Explanation}   \\
\hline
``label'' & Each entity is labeled as either 0 (benign), 1 (phishing), or 0.5 (unknown), reflecting its classification.\\
\hline
``predict\_label'' & The inferred label using our model initially assumes a value of $-1$, indicating an unknown classification.\\
\hline
``prior\_probability'' & A 2-element array containing probabilities for both benign and malicious classifications.\\
\hline
``msg\_sum'' & A 2-element array representing the sum of benign and malicious messages received from all its neighbors.\\
\hline
``msg\_nbr'' & A dictionary containing each neighbor along with the message they send. \\
\hline
\end{tabular}
\end{table}
\subsubsection{Loopy Belief Propagation}
%As a message passing algorithm, Loopy Belief Propagation (LBP) aims to gather information from neighboring nodes and utilize it to deduce the node label. In graphical models, messages can be computed using various algorithms such as the min-sum, max-sum, and sum-product methods. In our study, we employ the min-sum approach for message passing, which involves calculating messages by taking the minimum of incoming messages and adding a factor potential.
%
LBP is a message-passing algorithm that gathers information from neighboring nodes to infer node labels\cite{lbp}. It can be implemented using approaches like min-sum, max-sum, and sum-product. We will utilize the min-sum method, where messages are computed by taking the minimum of incoming messages and adding a factor potential.

%The LBP algorithm works by iteratively passing messages between neighboring entities in the graph until convergence is reached.
LBP iteratively passes messages between neighboring nodes in the graph until convergence. 
Consider an entity \( x \) belonging to the graph, where \( x \in X \) and \( N_{x} \) represent the set of its neighboring entities. Here, \( X \) encompasses both observed and hidden entities. In our study, observed entities represent benign and phishing URLs in the training dataset, while hidden entities denote URLs in the testing set and the extracted features. 

% The LBP algorithm proceeds as follows:

% 1. **Initialization**: Initialize messages for all edges in the graph. These messages typically start with some initial values.

% 2. **Message Passing**: Iterate over each entity \( x \) in the graph. For each entity:
%    - Compute outgoing messages from \( x \) to its neighboring entities \( N_{x} \), based on the current state of messages and any observed evidence.
%    - Update incoming messages received by \( x \) from its neighboring entities based on the outgoing messages from those neighbors.
%    - Repeat this process until convergence, where the messages stabilize and do not change significantly between iterations.

% 3. **Inference**: After convergence, compute the beliefs (posterior probabilities) for each entity \( x \) based on the final messages. These beliefs represent the inferred probabilities of the variables in the graphical model.

During the message-passing process, entity \( x \) sends messages to its hidden neighbors \( y \in N_{x} \) based on the messages it receives from other neighbors. Observed entities do not receive messages from neighbors as they do not require label prediction.
To compute messages \( msg_{x \hookrightarrow y}(l) \) from node \( x \) to node \( y \) regarding label \( l \in L \), where \( L = \{benign, phishing\} \).  The min-sum algorithm utilizes the following equation.
\begin{equation}
    \label{eq:msg}
    \setlength{\abovedisplayskip}{3pt}
    \setlength{\belowdisplayskip}{3pt}
    msg_{x \hookrightarrow y} (l) = \min_{\substack{l'}} [(1 - \phi_{x}(l')) + \psi_{xy}(l, l') + \displaystyle\sum_{k \in N_{x} \backslash y} msg_{k \hookrightarrow x} (l') ].
\end{equation}
Here, \( \phi_{x}(l') \) represents the prior probability that entity \( x \) is assigned label \( l' \).  For nodes in the training dataset, if the label is benign, the prior probability is set to $[1, 0]$, while for phishing URLs, it's $[0, 1]$. However, for nodes in the testing dataset and feature entities, their prior probability is set to $[0.5, 0.5]$ since they are considered hidden variables. Nonetheless, in our implementation, we integrate the prior probability obtained from the Random Forest Model.
Edge potential \( \psi_{xy} (l, l') \) represents the joint probability that node \( x \) has label \( l \) and node \( y \) has label \( l' \). In our study, we employed two types of edge potential assignment.
The term \( \sum_{k \in N_{x} \backslash y} msg_{k \hookrightarrow x} (l') \) sums all messages \( x \) has received from its neighbors, excluding \( y \).

When convergence is reached, indicating that the messages of all nodes have stabilized, we calculate the cost for each hidden node \( x \) having label \( l \) as follows 
\begin{equation}
\setlength{\abovedisplayskip}{3pt}
    \setlength{\belowdisplayskip}{3pt}
    \label{eq:cost}
    Cost(x, l) = (1 - \phi_{x}(l)) + \displaystyle\sum_{k \in N_{x}} msg_{k \hookrightarrow x} (l). 
\end{equation}

Following the min-sum algorithm, each node is assigned the label with the lowest cost, determined by comparing benign and malicious costs ${\arg\min}_{g}\displaystyle\sum_{x}Cost(x, g(x))$, where \( g(x) \) represents the cost of entity \( x \) with label \( L\in \{benign, phishing\} \). We derive two distinct cost values: benign cost and malicious cost, indicating the cost of \( x \) if labeled as benign or malicious, respectively. By comparing these costs, we determine the label of \( x \) based on the minimum cost. In this study, rather than directly comparing their values, we calculate their ratio and subsequently choose an appropriate classification threshold, as discussed in Convergence Strategy.\par

\subsubsection{Edge Potential}
\label{sec:edge}
We employed two types of edge potential assignments. The first is derived from Polonium technology \cite{big-oh}, which assigns an edge potential of \(0.5 + \varepsilon\) if \(x\) and \(y\) have different labels and \(0.5 - \varepsilon\) if they have the same label, as shown in Table \ref{table:edge_t1}. 
The second edge potential assignment proposed by \cite{ref}, incorporates the similarity among entities. As shown in Table \ref{table:edge_sim}, the edge potential is adjusted based on the similarity \(sim(x, y)\), calculated from the vector representations of nodes \(x\) and \(y\). 
In this study, we calculate similarity with Cosine similarity and Radial Basis Function (RBF) kernel as follows \cite{rbf}:
\begin{equation}
    \label{eq:similarity}
    \setlength{\abovedisplayskip}{3pt}
    \setlength{\belowdisplayskip}{3pt}
   sim(x, y) = \left\{ 
    \begin{array}{ll}
    \cos(x, y), & \text{Cosine};\\
    \exp\left(\frac{\lVert x - y \rVert ^2}{2\sigma ^ 2}\right), & \text{RBF kernel}.\\
    \end{array}
    \right.
\end{equation}

%Cosine similarity evaluates the similarity between two vectors based on the cosine of the angle between them. RBF kernel, also known as Gaussian similarity, assesses similarity based on the distance between points in the feature space. It's crucial to normalize the vectors when utilizing the RBF kernel, especially considering that different features may have different scales in our study. Failure to normalize the vectors can result in features with larger magnitudes exerting a disproportionate influence on the distance calculation, potentially leading to biased results.
Cosine similarity measures the similarity between two vectors by calculating the cosine of the angle between them, while the RBF kernel, or Gaussian similarity, assesses similarity based on the distance between points in feature space \cite{rbf}. It is essential to normalize vectors when using the RBF kernel to account for varying feature scales. Without normalization, features with larger magnitudes may disproportionately influence the distance calculation, leading to biased results.

The refined edge potential mechanism (Table \ref{table:edge_sim}) enhances the assignment of more sophisticated edge potentials compared to the default one. 
%In instances where two entities share the same label but demonstrate low similarity $sim(x, y)$, setting the edge potential excessively high (i.e., $1 - sim$) may result in erroneous classifications. To address this concern, a penalty term, denoted as $ths_+$, restricts the maximum edge potential values when entities share identical labels. Similarly, in cases where entities possess different labels, the penalty term ensures that the edge potential remains sufficiently large, even when the similarity between the entities is small. 
Specifically, when two entities share the same label but exhibit low similarity $sim(x, y)$, assigning a high edge potential (i.e., $1 - sim$) can lead to misclassification. To mitigate this, we introduce a penalty term $ths_+$, which limits the edge potential for entities with identical labels. Conversely, for entities with different labels, the penalty ensures the edge potential remains sufficiently large, even when similarity is low.
By considering both similarity and label relationships, this dynamic edge potential assignment enhances the model’s ability to differentiate between entities, improving classification accuracy.

% """Through Table\ref{table:edge_sim}, we can found that When the similarity between \(x\) and \(y\) is low, they appear dissimilar, indicating a higher cost should be assigned. Therefore, we apply a large penalty close to 1. Conversely, when the similarity is high and they have the same label, we calculate the edge potential using \(\min(ths_{+}, 1 - sim(x, y))\). In this case, since they share the same label, we aim to assign a smaller edge potential to the cost. Thus, \(ths_{+}\) should be a smaller value. However, if they have different labels, we use \(\max(ths_{-}, sim(x, y))\) to calculate the edge potential. Here, since they have different labels, we aim to assign a higher edge potential to the cost, meaning a larger \(ths_{-}\) is chosen. When similarity is low and x, y has same label, min(t+, 1-sim), limit sim to a minimum edge potential to ths+. 
% """

\begin{table}
\vspace{-0.2cm}
\centering
\caption{Edge potential assignment with $\varepsilon$. \label{table:edge_t1}}
\vspace{-0.2cm}
\begin{tabular}{ccc}
\hline
\multicolumn{1}{c|}{$\psi_{xy} (l, l')$}  &{Benign} &{Phishing}  \\
\hline
\multicolumn{1}{c|}{Benign}& $0.5 - \varepsilon$ & $0.5 + \varepsilon$\\
\multicolumn{1}{c|}{Phishing} & $0.5 + \varepsilon$ & $0.5 - \varepsilon$ \\
\hline
\end{tabular}
\end{table}

\begin{table}
\vspace{-0.2cm}
\centering
\caption{Edge potential assignment with similarity. \label{table:edge_sim}}
\vspace{-0.2cm}
\begin{tabular}{ccc}
\hline
\multicolumn{1}{c|}{$\psi_{xy} (l, l')$}  &{Benign} &{Phishing}  \\
\hline
\multicolumn{1}{c|}{Benign}& $\min(ths_{+}, 1 -sim(x, y)$ & $\max(ths_{-}, sim(x, y)$\\
\multicolumn{1}{c|}{Phishing} & $\max(ths_{-}, sim(x, y)$  & $\min(ths_{+}, 1 - sim(x, y)$ \\
\hline
\end{tabular}
\end{table}

\subsubsection{Entity Vectorization}

% Vectorization is a pivotal technique in Natural Language Processing (NLP), offering diverse approaches like word embedding-based methods such as word2vec, doc2vec, and network embedding-based methods like node2vec and deepwalk. In our study, we utilized word2vec to vectorize entities.

% Word2vec \cite{word2vec}, as a semantic embedding method, primarily focuses on modeling word semantics. While it can be extended to handle longer sequences of text, such as sentences or documents, by aggregating the word vectors within those sequences, it does not inherently capture the structural or sequential information present in longer text sequences. Therefore, directly generating vector representations for entities like domains, IP addresses, and nameservers using word2vec may not effectively capture their sequential nature.

% Instead, according to the locally linear embedding (LLE) \cite{LLE} technique, we can compute vector representations for these entities by equally weighting the combination of their neighbors. This sequential process begins with the domain, proceeds to IP addresses, and concludes with nameservers. This approach allows us to incorporate the sequential relationships between these entities in their vector representations, enhancing their effectiveness for downstream tasks.

Vectorization is a key technique in Natural Language Processing (NLP), with methods such as word2vec and network embedding techniques like node2vec and deepwalk being widely used\cite{ref}. In this study, we applied word2vec to vectorize entities.
Word2vec excels at capturing word semantics by generating vector representations of words \cite{word2vec}. However, while it can aggregate word vectors to handle longer text sequences, it does not inherently capture structural or sequential information, which is crucial for entities like domains, IP addresses, and nameservers.

To address this, we utilized the Locally Linear Embedding (LLE) technique \cite{LLE} to compute vector representations for these entities by weighting their neighboring nodes. This process begins with the domain, followed by IP addresses, and finally nameservers. By considering these sequential relationships, it enhances the effectiveness of the vector representations for downstream tasks.

\subsection{Convergence Strategy}
\label{sec:convergence}
% While Loopy Belief Propagation (LBP) is effective at handling graphs with known cycles, it encounters challenges if there are unknown cycles in the graph. In such cases, LBP will struggle to converge to a stable solution or may converge to incorrect beliefs due to the presence of conflicting information propagated through the graph.

% Upon explicit analysis, we identified the presence of unknown cycles in our graph, which exclusively involve hidden variables. Consequently, the messages associated with these hidden entities fail to converge. This phenomenon occurs because the hidden nodes within unknown cycles are influenced by their own messages during the message passing process, leading to continual increases or decreases in their message values over iterations.

While LBP is effective for graphs with known cycles, it faces difficulties when unknown cycles are present. In such cases, LBP may either fail to converge to a stable solution or converge to incorrect beliefs due to conflicting information being propagated through the graph.
In our analysis, we identified unknown cycles involving hidden variables within the graph. As a result, the messages associated with these hidden nodes fail to converge. This occurs because, during the message-passing process, hidden nodes in these cycles are influenced by their own messages, causing the message values to continually increase or decrease over iterations.

\begin{figure}
\setlength{\abovecaptionskip}{-0.2cm} 
\setlength{\belowcaptionskip}{-1cm}
\centering
\includegraphics[width=0.45\textwidth]{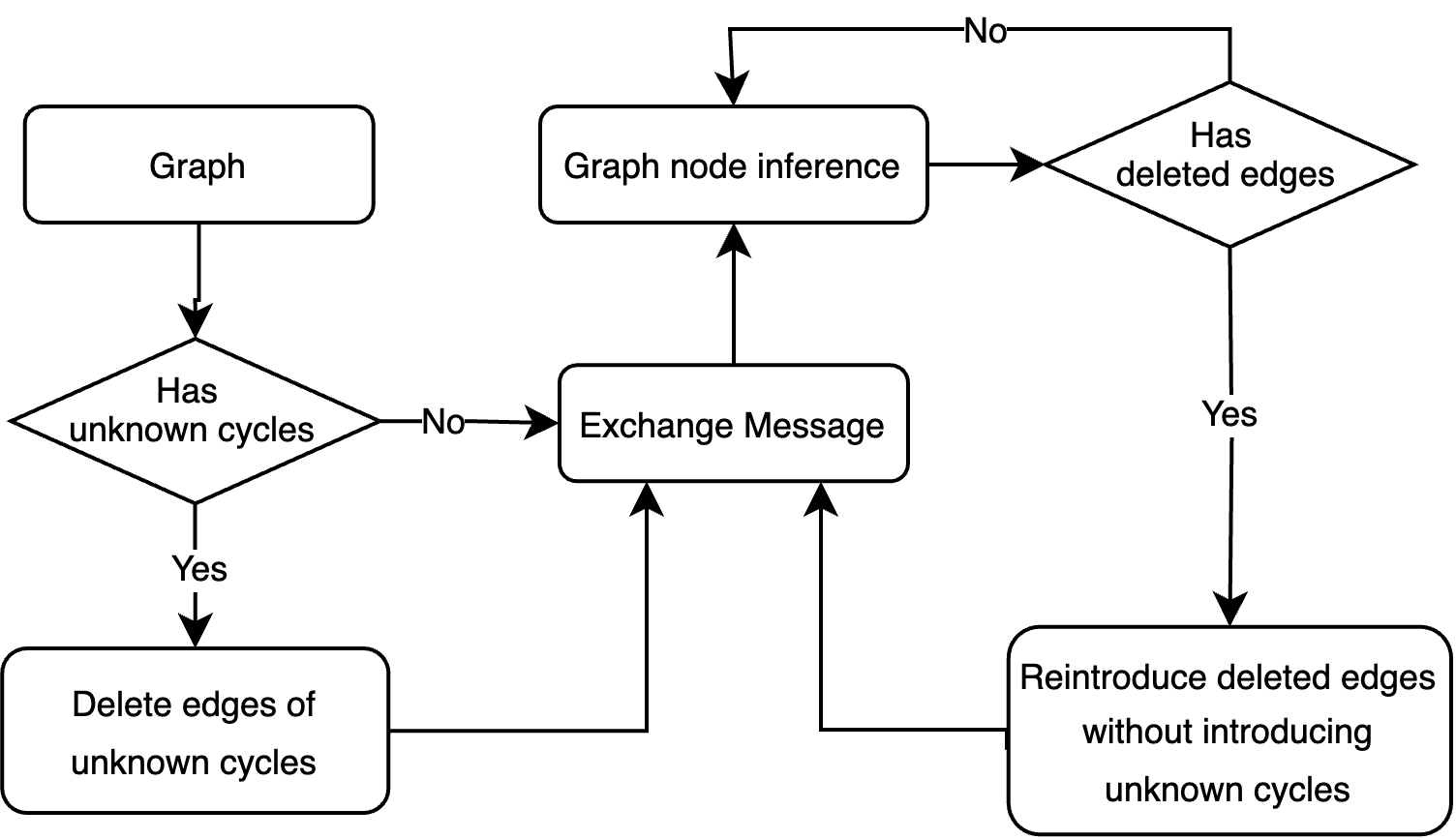}
\caption{The workflow of convergence approach.}
\label{fig:delete}
\end{figure}

To address this issue, we propose a new approach by deleting unknown cycles, as outlined in Fig.\ref{fig:delete}. Initially, we remove all edges of cycles consisting exclusively of hidden variables. Subsequently, we perform message passing and assign labels to variables with observed neighbors based on the current calculated cost values. Since the removed edges are between hidden variables, their influence is lesser than that of observed variables. Next, we reintroduce the deleted edges into the graph, ensuring they do not cause unknown cycles. We repeat these steps until there are no available deleted edges that can be added back to the graph. Consequently, we assign labels to the remaining unknown nodes based on the current messages.

In our implementation, nodes receive messages from their neighbors, which enhances efficiency. During each iteration, hidden nodes gather messages from all neighbors, storing the values temporarily. Once convergence is reached, these stored messages are updated in the graph, ensuring consistency and stability in the message-passing process, leading to accurate predictions and inference.

\section{Simulation Results}

In this study, we conduct a comprehensive evaluation of our novel phishing URL detection model. Additionally, we compare the performance of our new model against existing models as baselines. The source codes, data, and reproducibility information of our method are available at https://github.com/wenyeguo/LBP.

We utilize several datasets, including one from \cite{ref} and a newly collected dataset. The datasets consist of two types of URLs: benign (labeled 0) and phishing (labeled 1). 
The datasets exhibit an imbalance due to the transient nature of phishing URLs, which are often created for short-term malicious activities like phishing or malware distribution.
To address this imbalance, we fine-tune the classification threshold to allow the model to effectively capture the minority class (malicious URLs) while maintaining acceptable false positive rates.

%\subsection{Baselines}
%\label{sec:baselines}

\begin{figure}
\setlength{\abovecaptionskip}{-0.2cm} 
	 \setlength{\belowcaptionskip}{-1cm}
\centering
\includegraphics[width=0.45\textwidth]{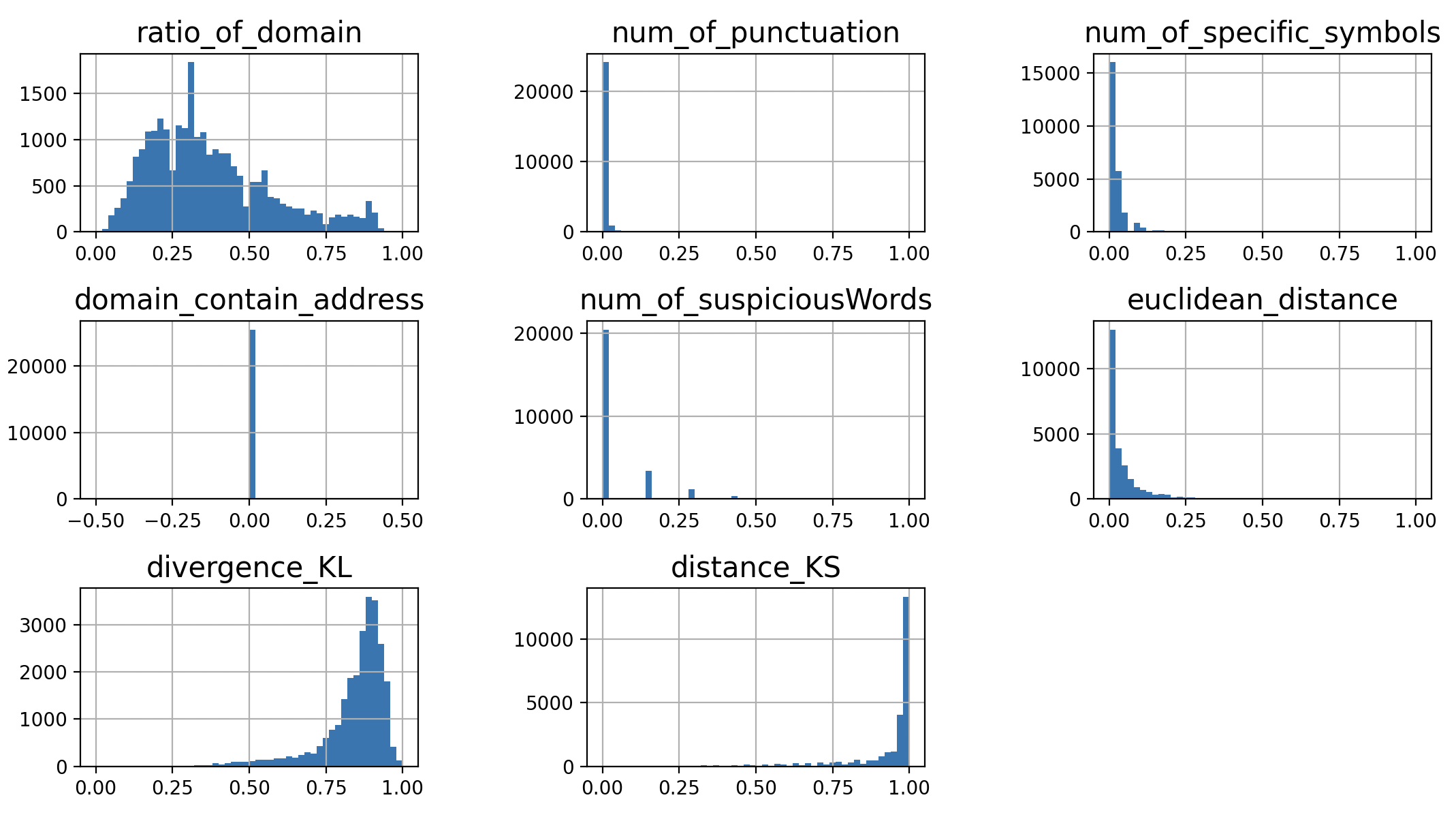}
\caption{Data distribution visualization.}
\label{fig:b_features}
\end{figure}

Our experiments are conducted on Apple M2 machines with 8-core CPUs and 8GB of memory, using Python 3.12.2, Networkx 3.3, Pandas 2.2.2, Numpy 1.26.4, and Scikit-learn 1.4.2.
The features utilized in the baseline models are sourced from \cite{baseline1, baseline2, baseline3, baseline4, baseline5}, focusing on characteristics of the URL strings. 
After visualizing the features extracted from our dataset (refer to Fig.\ref{fig:b_features}), we discover that the ``domain\_contain\_address'' feature lacks significance for training the machine learning model. Consequently, we will exclude this feature from the dataset before training the model.

We used three machine learning algorithms as baseline benchmarks to evaluate our model's effectiveness, i.e., Logistic Regression, Random Forest, and Naive Bayes.
Logistic Regression is a binary classification algorithm that uses the logistic function to map features to probabilities. It is widely used for tasks like anomaly detection.
Random Forest is an ensemble method that constructs multiple decision trees from random subsets of data to improve classification accuracy and reduce overfitting. It is robust and requires minimal tuning.
 Naive Bayes is a probabilistic classifier based on Bayes' theorem, assuming feature independence. It is efficient and well-suited for high-dimensional data. 

\begin{figure}
\setlength{\abovecaptionskip}{-0.2cm} 
	 \setlength{\belowcaptionskip}{-1cm}
\centering
\includegraphics[width=0.35\textwidth]{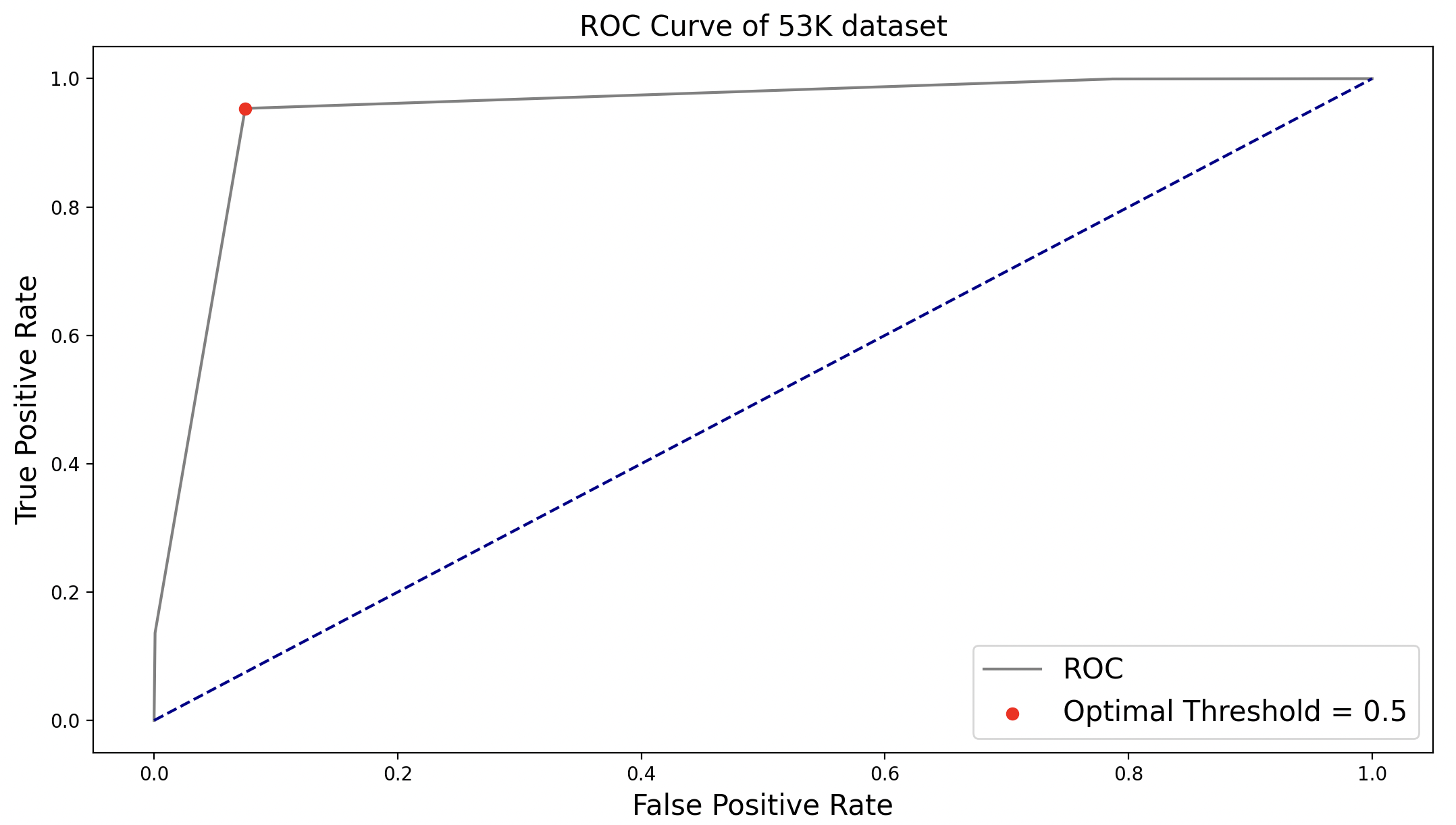}
\caption{ROC curve based on true positive rate (TPR) and false positive rate (FPR).}
\label{fig:roc}
\end{figure}

\subsection{Classification Threshold}
\label{sec:classification}
Adjusting the classification threshold is crucial for optimizing classifier performance, especially with imbalanced datasets. In our study, we use grid search and ROC curve analysis to find the optimal threshold that balances true positive and false positive rates.
We test thresholds ranging from 0 to 1 with increments of 0.1 and plot the ROC curve for our model on our dataset (Fig.\ref{fig:roc}). From this curve, we select the threshold that maximizes the F1 score, effectively balancing sensitivity and specificity. In our study, the optimal classification threshold is found to be 0.5.

\subsection{Evaluation Metrics}
%Given the imbalanced nature of our dataset, relying solely on accuracy may not offer a comprehensive evaluation of performance. Therefore, we utilize multiple metrics such as accuracy, precision, recall, and F1 score to assess the effectiveness of the classifier across various dimensions. We will delve into their definitions and calculation methods in the subsequent discussion.

% First, let's define the four different classes which serve as the basis for computing accuracy, precision, recall, and F1 score:
Given the imbalanced nature of our dataset, accuracy alone is insufficient for a comprehensive evaluation. Therefore, we assess classifier performance using multiple metrics: accuracy, precision, recall, and F1 score.
The following classes form the basis for these metrics:
(1) True Positive (TP): correctly predicting benign URLs as benign;
(2) False Positive (FP): incorrectly predicting malicious URLs as benign;
(3) True Negative (TN): correctly predicting malicious URLs as malicious;
(4) False Negative (FN): incorrectly predicting benign URLs as malicious.
Based on the above metrics, accuracy is defined to measure the proportion of correctly classified instances: $Accuracy = \frac{TP + TN}{TP + FP + TN + FN}$.
% \be
% Accuracy = \frac{TP + TN}{TP + FP + TN + FN}.
% \ee
Similarly, precision indicates how many predicted benign URLs are actually benign:
$Precision = \frac{TP}{TP + FP}$.
% \be
% Precision = \frac{TP}{TP + FP}.
% \ee
Recall reflects how many benign URLs are correctly predicted out of all actual benign URLs: $Recall = \frac{TP}{TP + FN}$.
% \be
% Recall = \frac{TP}{TP + FN}.
% \ee
The F1 score is defined as a weighted average of precision and recall:
$F1 \ Score = 2 \times \frac{Precision \times Recall}{Precision + Recall}$.
% \be
% F1 \ Score = 2 \times \frac{Precision \times Recall}{Precision + Recall}.
% \ee

\subsection{Results}
\begin{table}
\vspace{-0.2cm}
\centering
\caption{The performance of proposed method. \label{table:baselineVSnew}}
\vspace{-0.2cm}
\begin{tabular}{|ccccc|}
\hline
\textbf{Model}  &\textbf{Accuracy}  & \textbf{Precision} & \textbf{Recall} &\textbf{F1} \\
\hline
LogisticRegression & 79.57 & 78.91 & 73.57 & 76.14 \\
RandomForest & 87.53 & 87.76 & 83.5 & 85.58 \\
NaiveBayes & 75.86 & 70.14 & 79.28 & 74.43 \\
New Model & 94.1 & 94.12 & 95.35 & 94.73 \\
\hline
\end{tabular}
\end{table}

Among the three baseline models, Random Forest achieved the highest F1 score, as indicated in Table \ref{table:baselineVSnew}. Consequently, we opted to utilize Random Forest to derive the prior probability, which was then integrated into our new model.

\begin{figure}
\setlength{\abovecaptionskip}{-0.2cm} 
	 \setlength{\belowcaptionskip}{-1cm}
\centering
\includegraphics[width=0.45\textwidth]{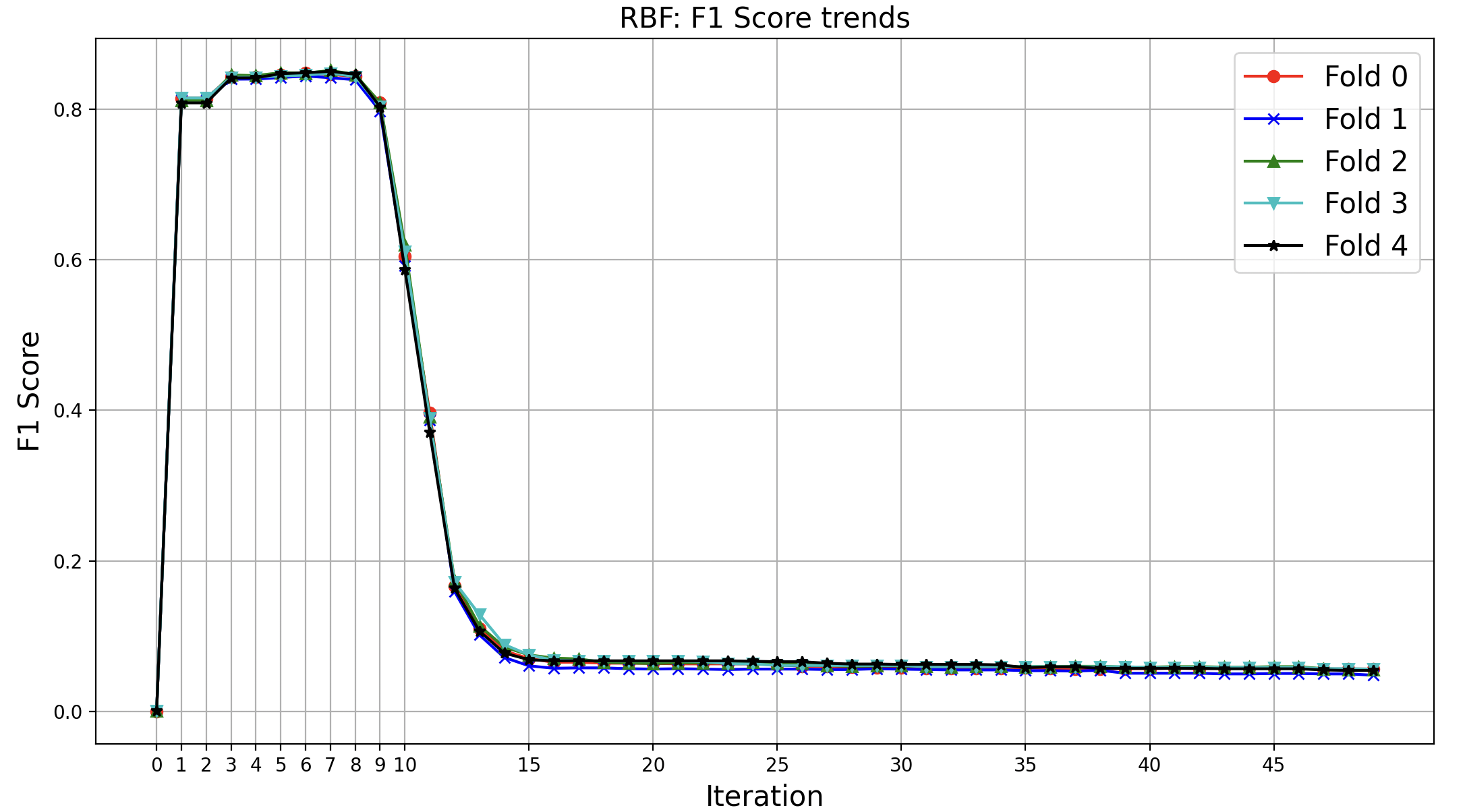}
\caption{F1 score trends over iterations.}
\label{fig:f1}
\end{figure}

We utilize a Dataset comprising 53,871 samples along with basic edge potential assignment, terminating after 6 iterations, to assess the impact of integrating prior probabilities derived from traditional models. Specifically, we compare utilizing the prior probability returned by Random Forest (RF) against a default probability of 0.5. As depicted in Table \ref{table:prior}, leveraging the prior probability from Random Forest yielded an 8.86\% enhancement in the F1 score. This highlights the significance of incorporating such adjustments into model construction. Furthermore, in comparison to the traditional machine learning Random Forest, our new model demonstrates an improvement of over 4\% in F1 score. This highlights the efficiency of the graph-based approach.

To determine the optimal number of message exchanges (\(k\)) for our original method, we conduct a comprehensive study. We iterate our model 100 times and plot the convergence nodes alongside the corresponding F1 scores, as illustrated in Fig.\ref{fig:f1}. Our analysis reveals that the convergence rate stabilizes after approximately 5 iterations. Additionally, we observe that the F1 score reaches its peak when \(k\) is between 5 to 7. Based on these findings, we select \(k = 6\) as the optimal value.%and Fig.\ref{fig:convergence}

\begin{table}
\vspace{-0.2cm}
\centering
\caption{The performance with different prior probabilities. \label{table:prior}}
\vspace{-0.2cm}
\begin{tabular}{|cccccc|}
\hline
\textbf{Model}  &\textbf{Prior Prob.}  &\textbf{Accuracy}  & \textbf{Precision} & \textbf{Recall} &\textbf{F1} \\
\hline
Graph & [0.5, 0.5] & 80.81 & 89.58 & 75.96 & 82.21\\
Graph & RF & 87.96 & 91.06 & 87.99 & 89.5 \\
RF & - & 87.53 & 87.76 & 83.5 & 85.58 \\
\hline
\end{tabular}
\end{table}

\begin{table}
\centering
\vspace{-0.2cm}
\caption{The performance with different convergence strategies. \label{table:convergence}}
\vspace{-0.2cm}
\begin{tabular}{|cccccc|}
\hline
\textbf{Method}  &\textbf{Accuracy}  & \textbf{Precision} & \textbf{Recall} &\textbf{F1} &\textbf{CVG}\\
\hline
k Times Iteration & 87.86 & 91.07 & 87.83 & 89.42 & 52\\
Delete Cycles & 91.46 & 90.87 & 94.11 & 92.47 & 98\\
\hline
\end{tabular}
\end{table}

To assess the effectiveness of our novel convergence (CVG) approaches, we employ the dataset with prior probability returned from Random Forest and basic edge potential assignment $0.5 \pm \varepsilon$. As depicted in Table \ref{table:convergence}, the new convergence method, ``delete unknown cycles'', exhibits an 88\% improvement in convergence rate and a 3.4\% increase in the F1 score compared to the scheme terminating after iterating k = 6 times. Additionally, a notable advantage of the new approach is its consistency across multiple runs on the same dataset, whereas iterating k times does not achieve consistency.

%\subsection{Edge potential}

\begin{table}
\centering
\vspace{-0.2cm}
\caption{The performance with different edge potential assignment\label{table:3edges}.}
\vspace{-0.2cm}
\begin{tabular}{|ccccc|}
\hline
\textbf{Method}  &\textbf{Accuracy}  & \textbf{Precision} & \textbf{Recall} &\textbf{F1}\\
\hline
$0.5 \pm \varepsilon$ & 91.46 & 90.87 & 94.11 & 92.47 \\
similarity only & 93.36 & 93.05 & 95.19 & 94.11 \\
similarity with thresholds & 94.1 & 94.12 & 95.35 & 94.73 \\
\hline
\end{tabular}
\end{table}

We evaluate three edge potential assignment methods, including one based solely on similarity values to emphasize the significance of the penalty. Assessments are carried out using the prior probability returned by Random Forest and the new convergence approach. Our results indicate that leveraging similarity values yields better performance than the $0.5 \pm \varepsilon$ method, as demonstrated in Table \ref{table:3edges}. This is because phishing URLs often contain similar strings, thus utilizing similarity aids in effectively identifying them. Through grid search, we determine that $ths_+ = 0.6$ and $ths_- = 1.0$ yield optimal performance.
Employing this, we evaluate our model across all datasets, as summarized in Table \ref{table:final_results}. Notably, our model consistently improves performance, even with highly imbalanced data in larger datasets. This is because larger datasets often provide a more comprehensive representation of underlying population or distribution, allowing model to learn more diverse patterns and relationships. Consequently, it achieves better generalization performance on unseen data.

\begin{table}[h]
\centering
\vspace{-0.2cm}
\caption{The performance of different datasets\label{table:final_results}.}
\vspace{-0.2cm}
\begin{tabular}{|ccccc|}
\hline
\textbf{Dataset}  &\textbf{Accuracy}  & \textbf{Precision} & \textbf{Recall} &\textbf{F1}\\
\hline
53,871 & 94.1 & 94.12 & 95.35 & 94.73 \\
100,000 & 95.56 & 96.07 & 98.18 & 97.12 \\
306,354 & 97.71 & 97.92 & 99.62 & 98.77 \\
\hline
\end{tabular}
\end{table}

% \begin{figure}[!t]
% \setlength{\abovedisplayskip}{3pt}
% 	\setlength{\belowdisplayskip}{3pt}
% \centering
%     \includegraphics[width=3 in]{systemmodel.eps}
%     \caption{The system model.} \label{Fig. 1}
%     \vspace{-0.5cm}
% \end{figure}

	\section{CONCLUSIONS}
	
	In this paper, we develop a phishing URL detection system leveraging Loopy Belief Propagation, edge potential assignment, convergence strategy, and hybrid features. Crafting an efficient feature list is crucial for enhancing the accuracy of the detection system. Therefore, we organize URL string-based features along with features associated with URLs.
Our experimental results show that computing edge potential using similarity leads to notable performance enhancements. Moreover, implementing the new convergence strategy, namely deleting unknown cycles, boosts the F1 score of the phishing detection model by more than 7\%. Additionally, compared to previous methods, the proposed approach demonstrates scalability, maintaining effectiveness even with large datasets.

% While the obtained results are satisfactory in terms of detection rate, enhancing the efficiency of the method is essential. Incorporating state-of-the-art learning software and technologies such as PyTorch could be beneficial. Additionally, given our relatively large dataset, employing parallel processing techniques could expedite training.\par

% Moreover, exploring new machine learning models capable of processing and analyzing data in real-time could enable rapid phishing attack detection as they unfold. This swift response can help mitigate the damage caused by such attacks.

	%\begin{thebibliography}{}
	\bibliographystyle{IEEEtran}	
	\bibliography{references.bib}
		
	%\end{thebibliography}	
\end{document}